\documentstyle{article}

\begin{document}

\title{
On integrable discretization of the inhomogeneous Ablowitz-Ladik model.
}

\author{
V. V. Konotop \\
Department of Physics and Center of Mathematical Sciences,\\ University
of Madeira, Pra\c{c}a do Munic\'{\i}pio, Funchal,
P-9000, Portugal.
}

\maketitle
\begin{abstract}

An integrable discretization of the inhomogeneous Ablowitz-Ladik model
with a linear force is introduced.  Conditions on parameters of the
discretization which are necessary for reproducing Bloch oscillations
are obtained.  In particular, it is shown that the step of the
discretization must be comensurable with the period of oscillations
imposed by the inhomogeneous force.  By proper choice of the step of the
discretization the period of oscillations of a soliton in the discrete
model can be made equal to an integer number of periods of oscillations
in the underline continuous-time lattice.

\end{abstract}

\section{Introduction}

During the last few years a great deal of attention has been paid to
integrable discretizations of nonlinear evolution equations.  The
interest is naturally justified by needs of the computational physics
\cite{AT,AL2}.  One of the purposes of integrable discretization
is the construction of a discrete analogue of a continuum model which
preserves main features of the last one.
This point becomes especially important when one deals
with inhomogeneous models.  In that case even "the first step" of the
discretization of a one-dimensional nonlinear evolution equation, i.e.
discrtization with respect to the spatial coordinate, may introduce
qualitatively new features into the dynamics.  So, for instance, in the
case of the inhomogeneous nonlinear Schr\"{o}dinger equation a
constant force, which  linearly depends on the spatial coordinate,
results only in the renormalization of the phase and
velocity of the one-soliton solution \cite{liu}
while the same force leads to oscillations of solitons
of the inhomogeneous Ablowitz-Ladik (AL) \cite{AL} model:
\begin{equation}
 \label{e1a}
 i\dot{q}_n +(1-q_nr_n)
(q_{n-1}+q_{n+1})+2\chi n q_n=0
\end{equation}
\begin{equation}
\label{e1b}
 -i\dot{r}_n +(1-q_nr_n) (r_{n-1}+r_{n+1})+2\chi n r_n=0
\end{equation}
(here $r_n=\pm \bar{q}_n$, $\chi$ is a real constant,
which from the physical point of view determines the strength of the
linear force, a dot stands for the derivative with respect to time, and
a bar stands for the complex conjugation) \cite{BLR,K,K1}.
Periodic dependence on time
is a property of any solution of (\ref{e1a}), (\ref{e1b}) and
it is caused by the discreteness. In the case of the one-soliton
solution which reads
\begin{equation}
\label{soliton}
q_n^{(s)}=-\bar{r}_n^{(s)}=\frac{\sinh(2w)}{\cosh[2nw-X(t)-X_0]}
e^{i[\Phi(t)-2n\chi(t-t_0)]}
\end{equation}
where
\begin{equation}
\label{solit2}
\Phi(t)=\frac 1\chi \cosh(2w)\sin[2\chi(t-t_0)]
\end{equation}
\begin{equation}
\label{solit3}
X(t)=\frac 1\chi \sinh(2w)\cos[2\chi(t-t_0)]
\end{equation}
$w$, $t_0$, and $X_0$ are real constants, the soliton dynamics
has deep analogy with the well known
Bloch oscillations of an electron in a lattice potential affected by a
constant electric field (due to this reason such behaviour
is referred to as Bloch oscillations \cite{K}). As it follows from
(\ref{solit3}) the period of oscillations is given by $\tau_0=\pi/\chi$.

The phenomenon of Bloch oscillations becomes especially interesting if
one looks for the possibility of integrable discretization of the
inhomogeneous AL model with respect to time.  Indeed, Bloch
oscillations are
characterised by the additional temporal scale, $\tau_0$. This scale is
determined by the strength of the force and must lead to some constrains
on the step of discretization.  Thus the purpose of the present
communication is to introduce integrable discretization of the model
(\ref{e1a}), (\ref{e1b}) and to obtain conditions on parameters of the
discretization which preserve the effect of Bloch
oscillations in the discrete scheme.

\section{Integrable discretization}

At $\chi=0$ system (\ref{e1a}), (\ref{e1b}) transforms to the
conventional AL model which
discretization is well known \cite{AL2,Suris}. In particular, it can be
achieved by using
the discrete analogue of the zero-curvature condition
\begin{equation}
\label{e2}
U(n,t+h)V(n,t)=V(n+1,t)U(n,t).
\end{equation}
In the case $\chi\neq0$ the same condition involving
$U$-matrix as follows
\begin{eqnarray}
\label{U}
U(n,t)=\left(\begin{array}{cc}
\lambda e^{-i\chi t} & q(n,t) \\
r(n,t) & e^{i\chi t}/\lambda
\end{array}\right)
\end{eqnarray}
and $V$-matrix having the elements
\begin{eqnarray}
\nonumber
V_{11}=i-h\alpha_0+h\alpha_1\left(\lambda^2e^{-i\chi
(2t+nh)}-A(n,t)\right)+
h\left(\frac{\alpha_2}{\lambda^2}e^{i\chi
(2t-nh)}-\delta_2q(n,t+h)r(n-1,t)\right)\Lambda(n,t)
\end{eqnarray}
\begin{eqnarray}
V_{12}=h\left(\alpha_1\lambda e^{-i\chi
(t+nh)}q(n,t)-\frac{\delta_1}{\lambda}e^{i\chi
(t+(n-1)h)}q(n-1,t+h)\right)+
\nonumber \\
h\left(\delta_2\lambda e^{-i\chi
(t-(n+1)h)}q(n,t+h)-\frac{\alpha_2}{\lambda}e^{i\chi
(t-nh)}q(n-1,t)\right)\Lambda(n,t)
\nonumber
\end{eqnarray}
\begin{eqnarray}
V_{21}=h\left(\alpha_1\lambda e^{-i\chi
(t+(n-1)h)}r(n-1,t+h)-\frac{\delta_1}{\lambda}e^{i\chi
(t+nh)}r(n,t)\right)+
\nonumber   \\
h\left(\delta_2\lambda e^{-i\chi
(t-nh)}r(n-1,t)-\frac{\alpha_2}{\lambda}e^{i\chi
(t-(n+1)h)}r(n,t+h)\right)\Lambda(n,t)
\nonumber
\end{eqnarray}
\begin{eqnarray}
\nonumber
V_{22}=i+h\delta_0-h\delta_1\left(e^{i\chi
(2t+nh)}/\lambda^2-D(n,t)\right)-
h\left(\delta_2\lambda^2e^{-i\chi
(2t-nh)}-\alpha_2q(n-1,t)r(n,t+h)\right)\Lambda(n,t)
\end{eqnarray}
where $\lambda$ is a spectral parameter, $\alpha_j$ and $\delta_j$ are
parameters and
$h$
($h>0$) is a step of the discretization,
 results in the system
\begin{eqnarray}
\label{e3}
ih^{-1}[q(n,t+h)-q(n,t)]=\delta_1q(n-1,t+h)e^{ih\chi (n-2)}-
\delta_0q(n,t+h)-
\nonumber \\
\alpha_0q(n,t)+\alpha_1q(n+1,t)e^{-ih\chi (n+1)}-
\delta_1 q(n,t+h)D(n,t)-
\alpha_1q(n,t) A(n+1,t)
\nonumber \\
\alpha_2q(n-1,t)[e^{-ih\chi(n+1)}-q(n,t+h)r(n,t+h)]\Lambda(n,t)+
\nonumber \\
+\delta_2q(n+1,t+h)[e^{ih\chi(n+2)}-q(n,t)r(n,t)]\Lambda(n+1,t),
\end{eqnarray}
\begin{eqnarray}
\label{e4}
-ih^{-1}[r(n,t+h)-r(n,t)]=\alpha_1r(n-1,t+h)e^{-ih\chi(n-2)}
-\alpha_0r(n,t+h)-
\nonumber \\
\delta_0r(n,t)+\delta_1r(n+1,t)e^{ih\chi (n+1)}-
\alpha_1 r(n,t+h)A(n,t)-\delta_1r(n,t) D(n+1,t) +
\nonumber \\
\delta_2r(n-1,t)[e^{ih\chi (n+1)}-q(n,t+h)r(n,t+h)]\Lambda(n,t)+
\nonumber \\
\alpha_2r(n+1,t+h)[e^{-ih\chi(n+2)}-q(n,t)r(n,t)]\Lambda(n+1,t)
\end{eqnarray}
\begin{eqnarray}
\label{e5}
\alpha_1\left[A(n+1,t)-A(n,t)e^{-ih\chi}\right]-h^{-1}(i-h\alpha_0)
\left(1-e^{-ih\chi}\right)=
\nonumber \\
\alpha_1\left[r(n,t)q(n+1,t)e^{-ih\chi (n+1)}-r(n-1,t+h)q(n,t+h)
e^{-ih\chi(n-1)}\right]+
\nonumber \\
\delta_2
r(n-1,t)q(n,t+h)\Lambda(n,t)
\left(e^{-ih\chi}-e^{ihn\chi}\right)-
\nonumber \\
\delta_2q(n+1,t+h)r(n,t)\left(1-e^{ih\chi (n+2)}\right)\Lambda(n+1,t)
\end{eqnarray}
\begin{eqnarray}
\label{e6}
\delta_1\left[D(n+1,t)-D(n,t)e^{ih\chi}\right]+
h^{-1}(i+h\delta_0)\left(1-e^{ih\chi}\right)=
\nonumber \\
\delta_1\left[r(n,t+h)q(n-1,t+h)e^{ih\chi(n-1)}-
r(n+1,t)q(n,t)e^{ih\chi(n+1)}\right]+
\nonumber \\
\alpha_2r(n,t+h)q(n-1,t)
\Lambda(n,t)\left(e^{ih\chi}-e^{-ih\chi n}\right)
+
\nonumber \\
\alpha_2r(n+1,t+h)q(n,t)\Lambda(n+1,t)\left(e^{-ih\chi(n+2)}-1\right)
\end{eqnarray}
\begin{equation}
\label{e6a}
\Lambda(n,t)[1-q(n,t+h)r(n,t+h)]=
\Lambda(n+1,t)[1-q(n,t)r(n,t)]
\end{equation}

Then the discrete analogue of the  AL model (\ref{e1a}), (\ref{e1b}) is
obtained from (\ref{e3})-(\ref{e6a}) by means of the reduction
\begin{equation}
\label{reduction}
r(n,t)=\pm \bar{q}(n,t)
\end{equation}
which requires the following relation among the parameters:
$\alpha_j=\bar{\delta}_j$.

In order to define
solutions of (\ref{e3})-(\ref{e6a}) one has to fix boundary conditions
for $r(n,t)$, $q(n,t)$, $A(n,t)$, and
$D(n,t)$. In what follows we deal only with the case of
 zero boundary conditions
\begin{equation}
\label{bound}
\lim_{n\to\pm\infty}q(n,t)=
\lim_{n\to\pm\infty}r(n,t)=0
\end{equation}
which allow existence of "bright" solitons. Hence it will be assumed
that $r(n,t)= -\bar{q}(n,t)$.
Respectively we have to require
\begin{equation}
\label{boundA}
\lim_{|n|\rightarrow\infty}A(n,t)=
A_0(n,t)=\frac{-i+h\alpha_0}{h\alpha_1}\left(e^{-ih\chi n}-1\right),
\end{equation}
\begin{equation}
\label{boundD}
\lim_{|n|\rightarrow\infty}D(n,t)=D_0(n,t)=
\frac{i+h\delta_0}{h\delta_1}\left(e^{ih\chi n}-1\right).
\end{equation}
Notice that (\ref{boundA}), (\ref{boundD}) transform
to the zero boundary conditions in the case
of the homogeneous AL model ($\chi=0$) \cite{Suris}.

Eqs. (\ref{e5}), (\ref{e6}) subject to (\ref{boundA}), (\ref{boundD})
allow one to express $A(n,t)$ and $D(n,t)$ through $q(n,t)$ and
$r(n,t)$:
\begin{equation}
\label{solA}
A(n,t)=A_0(n,t)+\alpha_1^{-1}\sum_{k=1}^{\infty}f_A(n-k,t)e^{-i\chi h k}
\end{equation}
\begin{equation}
\label{solD}
D(n,t)=D_0(n,t)+\delta_1^{-1}\sum_{k=1}^{\infty}f_D(n-k,t)e^{i\chi h k}
\end{equation}
Here $f_A(n,t)$ and $f_D(n,t)$ stand for the right hand sides of the
equations (\ref{e5}) and (\ref{e6}), correspondingly. It is to
emphasised that formulae (\ref{solA}) and (\ref{solD}) do not give yet
explicit solutions for $A(n,t)$ and $D(n,t)$ and after substitution into
(\ref{e3}), (\ref{e4}) represent a source of nonlocality of the discrete
scheme.

As it has been shown in \cite{K,K2} a convenient approach to treat
inhomogeneous discrete models is the use of the gauge transformation
which allows one to restrict the study only to the temporal
behaviour of the scattering data.
The gauge transformation in the discrete case is given by
\begin{equation}
\label{e3a}
\tilde{U}(n,t)=G(n+1,t)U(n,t)G^{-1}(n,t)
\end{equation}
\begin{equation}
\label{e3b}
\tilde{V}(n,t)=G(n,t+h)V(n,t)G^{-1}(n,t)
\end{equation}

By choosing $G(n,t)=\exp\{i\chi nt\sigma_3\}$, where $\sigma_3$ is the
Pauli
matrix one reduces $\tilde{U}(n,t)$ to the form which corresponds to
the $U$-matrix
of the underline homogeneous model (i.e. to the form which does not have
explicit dependence on time and can be obtained from (\ref{U}) by
the replacement $\exp (i\chi t)\mapsto 1$).
 Then the dependence of the transfer matrix
$T(t)$, associated with
$\tilde{U}(n,t)$, on the discrete time is governed by the equation
\begin{equation}
\label{time}
T(t+h)=V_hT(t)V_h^{-1}
\end{equation}
where $V_h$  is a diagonal matrix,
$V_h=$diag$(\theta_1(\lambda,t), \theta_2(\lambda,t))$, with the
elements
\begin{equation}
\label{theta1}
\theta_1(\lambda,t)=i-h\alpha_0+h\lambda^2\alpha_1e^{-2i\chi
t}+h\lambda^{-2}\alpha_2e^{2i\chi t}.
\end{equation}
\begin{equation}
\label{theta2}
\theta_2(\lambda,t)=i+h\delta_0-h\lambda^2\delta_2e^{-2i\chi
t}-h\lambda^{-2}\delta_1e^{2i\chi t}.
\end{equation}

Let us now
 assume that $t=mh$ and $m=0,1,...$ (it is straightforward to generalise
the results to the case $t=mh+t_0$ where $t_0$ is an arbitrary real
constant playing the role of initial moment of time). Then the element
$T^{(11)}$ of the matrix
$T(t)$ does
not depend on $m$ (or $t$) while for $T^{(12)}(t)\equiv b_m$ one obtains
\begin{equation}
\label{b1}
b_{m+1}=b_0 \prod_{n=0}^m\mu_n(\lambda)
\end{equation}
where
\begin{equation}
\label{b2}
\mu_m(\lambda)=\theta_1(\lambda,mh)/\theta_2(\lambda,mh).
\end{equation}

In the case of solitonic solutions (\ref{b1}) formally solves the
discrete Cauchy problem since
it defines dependence of the scattering data on time and the solution of
the eigenvalue problem for the matrix $U(n,0)$ is well known \cite{AL}.
Below we concentrate on some "physical" consequences of that
result.

\section{Bloch oscillations in the discrete model}

As it has been mentioned in the Introduction each step of
discretization can introduce
new features into the dynamics (even in cases of integrable models).
One of such features is the oscillatory behaviour of the solutions of
the AL model affected by the linear force (Bloch oscillations).  Now we
address to the question whether it is possible to preserve such
evolution subject to the discretization with respect to time.

To this end we take into account that periodic behaviour means
that there exists a positive integer $M$ such that
\begin{equation}
\label{c1}
\prod_{n=m+1}^{m+M}\mu_n(\lambda)=1
\end{equation}
for any $\lambda$ (which can be considered, say, inside the unit circle
on the complex plane) and any $m$. The period $\tau$ of oscillations
is then given by $\tau=Mh$ (evidently $M$ is considered to be
the smallest possible integer).

The discretzation of the homogeneous AL model is a three parametic one
(this, in particular, allows one to represent it in a form of a product
of the local maps \cite{Suris}). In the inhomogeneous case
the imposed conditions lead to constrains on the
parameters.    To find them we first consider the limit
$|\lambda|\to 0$ (or $|\lambda|\to\infty$). Then from (\ref{theta1}),
(\ref{theta2}), (\ref{b2}), and (\ref{c1}) one finds that there must
exist relations
\begin{equation}
\label{c2}
a_1=-a_2   \hspace{1 cm}
\phi_1^{(l)}+\phi_2^{(l)}=2\pi\frac{l}{M},
 \hspace{1 cm} l=0,1,...,M-1
\end{equation}
where
  $a_{1,2}$ and $\phi_{1,2}$
are real prameters connected to $\alpha_{1,2}$,
 $\alpha_{1,2}=a_{1,2}\exp({\phi_{1,2}})$,
  and the upper index has
been
attributed to the "quantized" phases. Next, the independence of
(\ref{c1}) on
$m$ implies
$\mu_m=\mu_{m+M}$ which means that
\begin{equation}
\label{chi}
\chi h_{\tilde{l}} M=\pi \tilde{l}
\end{equation}
where $\tilde{l}$ is a positive integer.
In other words the step of the discretization is not arbitrary
(the subindex $\tilde{l}$ is introduced to label different discrete
values of $h$).
Physical sense of the last requirement is quite transparent.
Recalling that the
period of Bloch oscillations in the continuous-time model is given by
$\tau_0=\pi/\chi$, one concludes that (\ref{c2}) means that the period
of the Bloch
oscillations in the discretized model, $\tau$, is $\tilde{l}$ times
bigger than the period
$\tau_0$: $\tau=\tilde{l}\tau_0$ and the number $\tilde{l}$ is related
to the chosen step of discretization $h$.
 On the other hand rewriting (\ref{chi})
as $h_{\tilde{l}}=\tilde{l}\tau_0/M$ one can interpret it as a condition
for the
discretization step to be comensurable with the period of oscillations.
 As it is evident, for the direct coincidence of the
result obtained on the discrete lattice with its continuum
counterpart one must let $\tilde{l}=1$. Below we concentrate on this
case.
Then $\mu_n(\lambda)$ takes the form
\begin{equation}
\label{mu1}
\mu_n(\lambda)=\frac{1+ae^{i(\Gamma_l+\gamma_{l,n})}
\lambda^2-ae^{i(\Gamma_l-\gamma_{l,n})}\lambda^{-2}}{
-1-ae^{-i(\Gamma_l+\gamma_{l,n})}
\lambda^{-2}+ae^{-i(\Gamma_l-\gamma_{l,n})}\lambda^2}e^{2i\phi_0}
\end{equation}
where
\[
\Gamma_l=\frac{l}{M}\pi-\phi_0, \hspace{1
cm}
\gamma_{l,n}=\frac
12
\left(\phi_1^{(l)}-\phi_2^{(l)}\right)-\frac{2\pi n}{M}
\]
$a=|h \alpha_1/(i-h\alpha_0)|$ and
$\phi_0=\arg (i-h\alpha_0)$.

Now we consider the unit circle where
$\lambda^2=\exp(i\psi)$ ($\psi$ being real).
 Then one can find two possibilities to satisfy the requirement
(\ref{c1}).
One simplest solution  corresponds to $M$ even and
$\phi_0=\pi l/M+\pi/2+\pi p$ ($p$ is an
integer). Then $\mu_n(\lambda)=\exp\{2\pi i(l/M-1/2)\}$. By the direct
algebra one ensures that this is the degenerated case, when the limiting
transition $h\to 0$ results in a trivial linear equation instead of the
AL model.
A nontrivial and physically relevant solution corresponds to the case
when $M=4N$ ($N$ is an integer) and $\phi_0=\phi_{l,p}= \pi
l/M+\pi p$  (in that case $\mu_n\mu_{n+N}\mu_{n+2N}\mu_{n+3N}=1$).
Then $\mu_n(\lambda)$ which determins evolution of the one-soliton
solution
associated with the eigenvalue $\lambda_1=\exp(-w+i\theta)$ is given by
\begin{equation}
\label{mu2}
\mu_n(\lambda_1)=-\frac{1-2(-1)^pa\sinh[2w-i(\gamma_{l,n}+2\theta)]}
{1+2(-1)^pa\sinh[2w-i(\gamma_{l,n}+2\theta)]}\exp\left(2\pi
i\frac{l}{M}\right)
\end{equation}

Let us illustrate the discrete-time dynamics on example of the
one-soliton solution. We assume that (\ref{c2}) holds. For the sake of
simplicity we let $\phi_1=-\phi_2=\pi/2$ and  $\phi_0=0$. Then the
one-soliton solution of (\ref{e3}), (\ref{e4}) can be written down in
the form (recall $t=mh$)
\begin{equation}
\label{dissol}
q^{(s)}(n,m)=-\bar{r}^{(s)}(n,m)=\frac{\sinh(2w)}{\cosh(2nw-X_m)}
e^{i\Phi(n,m)}
\end{equation}
where
\begin{equation}
\label{dissol2}
\Phi(n,m)=\sum_{k=0}^{m-1}\arctan\left[\frac{4a \cos
\left(2\chi kh-2\theta\right)\cosh(2w)}
{1+2a^2\cos\left(4\chi kh-4\theta
\right)+4a^2\cosh(4w)}\right]-2i\chi nmh
\end{equation}
\begin{equation}
\label{dissol3}
X_m=\frac 12 \sum_{k=0}^{m-1}\ln\left[
\frac{2a^2\left(\cosh(4w)+\cos\left(4\chi kh -4\theta \right)\right)
+1-4a\sinh(2w)\sin\left(2\chi kh-2\theta\right)}
{2a^2\left(\cosh(4w)+\cos\left(4\chi kh -4\theta\right)\right)
+1+4a\sinh(2w)\sin\left(2\chi kh-2\theta \right)}\right]
\end{equation}
Comparing these formulae with (\ref{solit2}), (\ref{solit3}) one can
see that choosing $a=h/2$ the last ones are obtained by the limiting
transition $h\to 0$.

To be more specific we concentrate on $X_m$ which describes evolution of
the centre of the soliton. To this end we represent
$h=\pi\xi/(M\chi)$ (the so chosen step
of the discretization satisfies (\ref{chi}) when $\xi$ is integer).
The results are summarized in Fig. 1  for three
values of the parameter $\xi$ displaying different
situations: (a) When $\xi=1$ the discrete model exactly reproduces Bloch
oscillations of the continuous-time model. (b) At $\xi=\sqrt{3}$ the
evolution of the discrete model is not periodic (notice that lines in
the figure are used for the sake of convenience of presentation: the
truth trajectories are sets of points).
 However, by considering analytic continuation of the solution the
periodicity can be considered between discrete time steps
\cite{AL2}. (c) At $\xi=2$ the period of soliton oscillations is two
times more than the period of Bloch oscillation of the AL model which is
obtained by the limiting transition $h\to 0$, $M\to \infty$ with
$hM=\pi/\chi$ (it is to be
mentioned that the minima of the curve (c) about $h=48$ and $h=145$ are
not numerical zeros).

Author acknowledges support from FEDER and Program
PRAXIS XXI, grant N$^0$ PRAXIS/2/2.1/FIS/176/94.


\vspace{1 true cm}
\begin{center}
Figure caption
\end{center}

Trajectory of the soliton centre corresponding to different steps of
discrete time (a) $h=1/97$, (b) $h=\sqrt{3}/97$, and (c) $h=2/97$. Other
parameters are as follows: $M=97$, $w=0.5$, $\chi=\pi$, $a=0.1$.

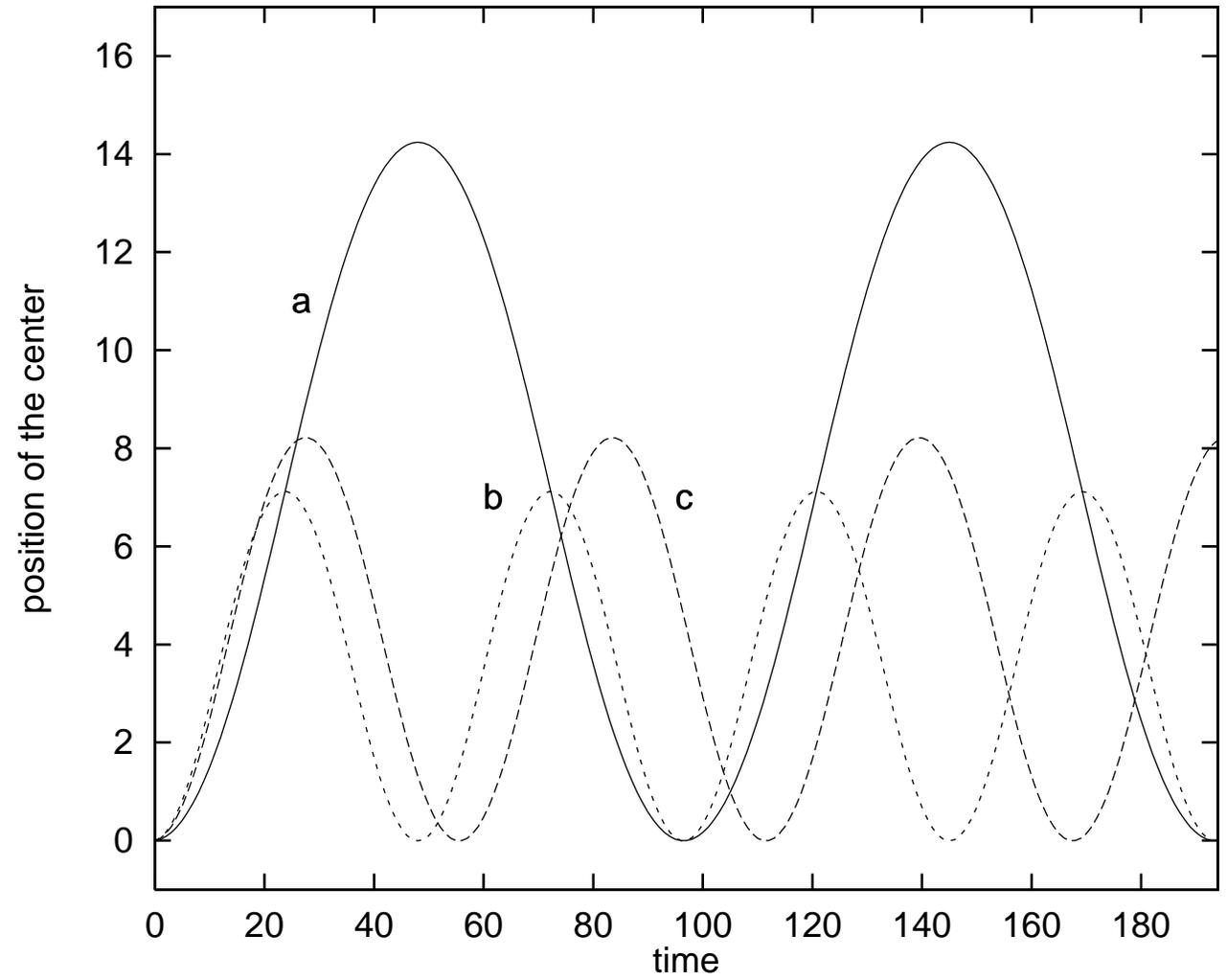
\end{document}